%% file: paper.tex
\documentclass{eptcs}
\usepackage{graphicx,url}
\usepackage[T1]{fontenc} 
\usepackage{xcolor}
\usepackage{amsmath}
\usepackage{amssymb}
\usepackage{amsthm}
\usepackage{wrapfig}

\usepackage{tikz}
\usetikzlibrary{arrows}

\usepackage{listings}
\usepackage{ded}
\usepackage{twelf-math}
\usepackage[nobook]{theorems}
\usepackage{basics}
\usepackage{local}
\def\titlerunning{Representing Isabelle in LF}
\def\authorrunning{F.~Rabe}

\begin{document}

\title{Representing Isabelle in LF}
\author{Florian Rabe
\institute{Jacobs University Bremen}
\email{f.rabe@jacobs-university.de}
}

\maketitle{}

\begin{abstract}
LF has been designed and successfully used as a meta-logical framework to represent and reason about object logics. Here we design a representation of the Isabelle logical framework in LF using the recently introduced module system for LF. The major novelty of our approach is that we can naturally represent the advanced Isabelle features of type classes and locales.

Our representation of type classes relies on a feature so far lacking in the LF module system: morphism variables and abstraction over them. While conservative over the present system in terms of expressivity, this feature is needed for a representation of type classes that preserves the modular structure. Therefore, we also design the necessary extension of the LF module system.
\end{abstract}

\section{Introduction}\label{sec:intro}
   \input{intro}
   
\section{Preliminaries}\label{sec:isa:prelim}
  \subsection{Isabelle}\label{sec:isa:prelim:isa}
    \input{isabelle}
  \subsection{LF}\label{sec:isa:prelim:lf}
    \input{lf}

\section{Morphism Variables in LF}\label{sec:isa:classes}
  \input{classes}
 
\section{Representing Isabelle in LF}\label{sec:isa:isa}
  \input{isa_lf}

\section{Conclusion}\label{sec:conc}
  \input{conc}
 
\bibliographystyle{plain}
\input{paper.bbl}

\end{document}

%% file: intro.tex
Both Isabelle and LF were developed at roughly the same time to provide formal proof theoretic frameworks in which object logics can be defined and studied. Both use the Curry-Howard correspondence to represent the proofs of the object logic as terms of the meta-logic.

Isabelle \cite{isabelle1,isabelle} is based on intuitionistic higher-order logic \cite{churchtypes} with shallow polymorphism and was designed as a generic LCF-style interactive theorem prover. LF \cite{lf} is the corner of the $\lambda$-cube \cite{lambdacube} that extends simple type theory with dependent function types and is inspired by the judgments-as-types methodology \cite{martinlof}. We will work with the Twelf implementation of LF \cite{twelf}.

It is straightforward to represent Isabelle's underlying logic as an object logic of LF (see, e.g., \cite{lf}). However, Isabelle provides a number of advanced features that go beyond the base logic and that cannot be easily represented in other systems. These include in particular a module system \cite{isabelle_locales,isabelle_classes} and a structured proof language \cite{isar}.

Recently, we gave a module system for LF in \cite{RS:twelfmod:09}. We wanted to choose primitive notions that are so simple that they admit a completely formal semantics. While such formal semantics are commonplace for type theories -- in the form of inference systems -- they quickly get very complex for module systems on top of type theories. At the same time these primitives should be expressive enough to admit natural representations of modular design patterns. Here by ``natural'', we mean that we are willing to accept lossy (in the sense of being non-invertible) encodings of modular specifications as long as their modular structure of sharing and reuse is preserved.

In this paper we give such a representation of the Isabelle module system in the LF module system. The main idea of the encoding is that all modules of Isabelle (theories, locales, type classes) are represented as LF signatures, and that all relations between Isabelle modules (imports, sublocales, interpretations, subclasses, instantiations) are represented as LF signature morphisms.

Thus, our contribution is two-fold. Firstly, we validate the design of the LF module system by showing that it provides just the right primitives needed to represent the Isabelle module system. Actually, before arriving at that conclusion we identify one feature that we have to add to the LF module system: abstraction over morphisms. And secondly, we show how LF can be used as a concise intermediate language in order to translate Isabelle libraries to other systems. Moreover, for researchers familiar with LF but not with Isabelle, this paper can complement the Isabelle documentation with an LF-based perspective on the foundations of Isabelle. However, an implementation of our representation must remain future work.

In Sect.~\ref{sec:isa:prelim}, we will repeat the basics of Isabelle and LF to make the paper self-contained. In Sect.~\ref{sec:isa:classes}, we extend the LF module system with abstraction over morphisms. Then we give our representation in Sect.~\ref{sec:isa:isa}.

%% file: isabelle.tex
Isabelle is a mature and widely used system, which has led to a rich ontology of Isabelle declarations. We will only consider the core and module system declarations in this paper. And even among those, we will restrict attention to a proper subset of Isabelle's power.

For the purposes of this paper, we make some minor adjustments for simplicity and consider Isabelle's language to be generated by the grammar in Fig.~\ref{fig:isa}. Here $|$ and $^*$ denote alternative and repetition, and we use special fonts for \nont{nonterminals} and \keyw{keywords}.

\begin{figure}[htb]
\begin{center}
\begin{tabular}{|lcl|}
\hline
\nont{theory}  & ::= & \keyw{theory} \nont{name} \keyw{imports} \nont{name}$^*$ \keyw{begin} \nont{thycont} \keyw{end} \\
\nont{thycont} & ::= & (\nont{locale} | \nont{sublocale} | \nont{interpretation} | \\
               &  |  &  \nont{class}  | \nont{instantiation} | \nont{thysymbol})$^*$ \\
\nont{locale}  & ::= & \keyw{locale} \nont{name} =
                       (\nont{name} : \nont{instance})$^*$ for \nont{locsymbol}$^*$ + \nont{locsymbol}$^*$ \\
\nont{sublocale}& ::= & \keyw{sublocale} \nont{name} $<$ \nont{instance} \nonti{proof}$^*$\\
\nont{interpretation}& ::= & \keyw{interpretation} \nont{instance} \nonti{proof}$^*$ \\
\nont{instance} & ::= & \nont{name} \keyw{where} \nont{namedinst}$^*$ \\
\nont{class}    & ::= & \keyw{class} \nont{name} = \nont{name}$^*$ + \nont{locsymbol}$^*$ \\
\nont{instantiation}& ::= & \keyw{instantiation} \nont{type} :: (\nont{name}$^*$)\nont{name}
                             \keyw{begin} \nont{locsymbol}$^*$ \nont{proof}$^*$ \keyw{end} \\
\hline
\nont{thysymbol} & ::= & \keyw{consts} \nont{con} | \keyw{defs} \nont{def} | \keyw{axioms} \nont{ax}
                      | \keyw{lemma} \nont{lem} \\
                 & |   & \keyw{typedecl} \nont{typedecl} | \keyw{types} \nont{types} \\
\nont{locysymbol}& ::= & \keyw{fixes} \nont{con} | \keyw{defines} \nont{def}
                      | \keyw{assumes} \nont{ax} | \keyw{lemma} \nont{lem} \\
\nont{con}      & ::= & \nont{name} :: \nonti{type} \\
\nont{def}      & ::= & \nont{name} : \nont{name} \nonti{var}$^*$ $\equiv$ \nonti{term} \\
\nont{ax}       & ::= & \nont{name} : \nonti{Prop} \\
\nont{lem   }                 & ::= & \nont{name} : \nonti{Prop} \nonti{proof} \\
\nont{typedecl} & ::= & (\nonti{var}$^*$) \nont{name} \\
\nont{types}    & ::= & (\nonti{var}$^*$) \nont{name} = \nonti{type} \\
\nont{namedinst} & ::= & \keyw{name} = \nonti{term} \\
\hline
\nonti{type} & ::= & \nonti{var} :: \nonti{name} | \nont{name} | $(\nonti{type},\ldots,\nonti{type})\;\nont{name}$ | $\nonti{type}\func\nonti{type}$ | $\prop$ \\
\nonti{term} & ::= & \nonti{var} | \nont{name} | $\nont{name}\;\nonti{term}^*$ |
  $\lambda (\nonti{var}::\nonti{type})^*.\nonti{term}$ \\
\nonti{Prop} & ::= &  $\nonti{Prop}\iimpl \nonti{Prop}$ | $\bigwedge(\nonti{var}::\nonti{type})^*.\nonti{Prop}$
                   |  $\nonti{term}\equiv\nonti{term}$ \\
\nonti{proof}& ::= & a primitive Pure inference as defined in \cite[p. 7]{isabellemanual} \\
\nont{name}, \nonti{var} &  ::= & identifier \\
\hline
\end{tabular}
\end{center}
\caption{Simplified Isabelle Grammar}\label{fig:isa}
\end{figure}

A \nont{theory} is a named group of declarations. Theories may use \keyw{imports} to import other theories, which yields a simple module system. Within theories, \nont{locale} and type \nont{class} declarations provide further sources of modularity. Theories, locales, and type classes may be related using a number of declarations as described below.
\medskip

The \emph{core declarations} occurring in theories (\nont{thysymbol}) and locales (\nont{locsymbol}) are quite similar. \nont{consts} and \nont{fixes} declare typed constants $c::\tau$. \nont{defs} and \nont{defines} declare definitions for a constant $f$ taking $n$ arguments as $f\_def : \; f\;x_1\;\ldots x_n \equiv t$ where $t$ is a term in the variables $x_i$. \nont{axioms} and \nont{assumes} declare named axioms $a$ asserting a proposition $\phi$ as $a:\phi$. \nont{lemma} declares a named lemma $l$ asserting $\phi$ with proof $P$ as $l:\phi\;P$.

Furthermore, in theories, \nont{typedecl} declares $n$-ary type operators $t$ as $(\alpha_1,\ldots,\alpha_n)\;t$, and similarly \nont{types} declares an abbreviation $t$ for a type $\tau$ in the variables $\alpha_i$ as $(\alpha_1,\ldots,\alpha_n)\;t=\tau$. Locales do not contain type declarations. However, they may declare new types indirectly by declaring constants whose types have free type variables, e.g., $\circ:\alpha\func\alpha\func\alpha$ in a locale for groups. References to these types are made indirectly using type inference, e.g., if there is another constant $e:\beta$, then an axiom $x\circ e=x$ enforces that $\alpha$ and $\beta$ refer to the same type.

\medskip
The constant declarations within a locale serve as parameters that can be instantiated. The intuition is that a locale \nont{instance} $loc$ \keyw{where} $\sigma$ takes the locale with name $loc$ and translates it into a new context (which can be a theory or another locale). Here $\sigma$ is a list of parameter instantiations (\nont{namedinst}) of the form $c=t$ instantiating the parameter $c$ of $loc$ with the term $t$ in that new context.

Locale instances are used in two places. Firstly, locale declarations may contain a list of instances used to inherit from other locales. In a locale declaration
 \begin{center}
 \keyw{locale} $loc$ = $ins_1:loc_1$ \keyw{where} $\sigma_1$ \;\ldots \; $ins_n:loc_n$ \keyw{where} $\sigma_n$
 \keyw{for} $\Sigma$ + $\Sigma'$
 \end{center}
the new locale $loc$ inherits via $n$ named instances: Instance $ins_i$ inherits from the locale $loc_i$ via the list of parameter instantiations $\sigma_i$. $\Sigma$ and $\Sigma'$ declare the core declarations of the locale.

The set of constant declarations of the locale is defined as follows: (i) The declarations in $\Sigma$ logically precede the instances, i.e., are available in $\sigma_i$ and $\Sigma'$. (ii) A copy of the declarations of each $loc_i$ translated by $\sigma_i$ is available in each $\sigma_j$ for $j>i$ and in $\Sigma'$; the names $ins_i$ serve as qualifiers to resolve name clashes if two declarations of the same name are present. (iii) The declarations in $\Sigma'$ are only available in $\Sigma'$.

The $\sigma_i$ do not have to instantiate all parameters of $loc_i$ -- parameters that are not instantiated become parameters of $loc$. Thus, the parameters of $loc$ consist of the not-instantiated parameters of the $loc_i$ and the constants declared in $\Sigma$ and $\Sigma'$.

Secondly, a declaration \keyw{sublocale} $loc'<loc$ \keyw{where} $\sigma\; \pi$ postulates a translation from $loc$ to $loc'$, which maps the parameters of $loc$ according to $\sigma$. The axioms and definitions of $loc$ induce proof obligations over $loc'$ that must be discharged by giving a list $\pi$ of proofs. If all proof obligations are discharged, all theorems about $loc$ can be translated to yield theorems about $loc'$, and Isabelle does that automatically. A locale \keyw{interpretation} is very similar to a \keyw{sublocale}. The difference is that all $loc$ expressions are translated into the current theory rather than into a second locale.
\medskip

The concepts of locales and \emph{type classes} have recently been aligned \cite{isabelle_classes}, and in particular type classes are also locales. But the syntax still reflects their different use cases. A type class is a locale inheriting only from other type classes and only without parameter instantiations. Thus, the locale syntax can be simplified to \keyw{class} $C$ = $C_1 \ldots C_n + \Sigma$ where $C$ inherits from the $C_i$. All declarations in $\Sigma$ may refer to at most one type variable, which can be assumed to be of the form $\alpha::C$. The intuition is that $\Sigma$ provides operations $c_1,\ldots,c_n$ that are polymorphic in the parametric type $\alpha$ and axioms about them.

An instance of a type class is a tuple $(\tau,c_1\_def,\ldots,c_n\_def)$ where $\tau$ is a type and $c_i\_def$ is a definition for $c_i$ at the type $\tau$. Because every $c_i$ can only have one definition per type, the definitions can be inferred from the context and be dropped from the notation; then a type class can be seen as a unary predicate on types $\tau$. Type class instantiations are of the form
\[\keyw{instantiation}\;t :: (C_1,\ldots,C_n)C\;\keyw{begin} \;\Sigma\;\pi\; \keyw{end}\]
 where $t$ is an $n$-ary type operator, i.e., a type with $n$ free type variables $\alpha_i$. $\Sigma$ contains the definitions for the operations of $C$ at the type $(\alpha_1,\ldots,\alpha_n)t$ in terms of the operations of the instances $\alpha_i::C_i$. This creates proof obligations for the axioms of $C$, and we assume that all the needed proofs are provided as a list $\pi$. The semantics is that if $\tau_i::C_i$ are type class instances, then so is $(\tau_1,\ldots\tau_n)t::C$. Note that this includes base types for $n=0$.

\begin{example}\label{ex:isa:isa}
The following sketches two type classes for orderings and semilattices with universe $\alpha$, ordering $\leq$, and infimum $\sqcap$ (where we omit inferable types and write $\cdot$ for empty lists):
\[\begin{array}{llcl}
\keyw{class} & order  &=& \cdot\; +\; \leq::\alpha\func\alpha\func\prop \\
\keyw{class} & semlat &=& order\; +\; \sqcap :: \alpha\func\alpha\func\alpha \\
\keyw{locale}& lat    &=& inf:semlat \;\keyw{where}\; \cdot \\
              &&&         sup:semlat \;\keyw{where}\; \leq=\lambda x\lambda y.\;y\;inf.\leq\; x \;\keyw{for}\; \cdot\;+\;\cdot \\
\end{array}\]
Here the omitted axioms in $semlat$ would enforce that the type variables $\alpha$ in the types of $\leq$ and $\sqcap$ refer to the same type.
Then a locale for lattices is obtained by using two named instances of a semilattice where the second one flips the ordering. The parameters of $lat$ are $inf.\leq$ (the ordering), $inf.\sqcap$ (the infimum), and $\sup.\sqcap$ (the supremum), but not $\sup.\leq$, which is instantiated.
\end{example}
\medskip


Finally the \emph{inner syntax} for \nont{term}s, \nont{type}s, \nont{prop}ositions, and \nont{proof} terms -- also called the Pure language -- is given by an intuitionistic higher-order logic with shallow polymorphism. Types are formed from type variables $\alpha::C$ for type classes $C$, base types, type operator applications, function types, and the base type $\prop$ of propositions. Type class instances of the form $\tau::C$ are formed from type variables $\alpha::C$ and type operator applications $(\tau_1,\ldots,\tau_n)t$ for a corresponding instantiation $t::(C_1,\ldots,C_n)C$ and type class instances $\tau_i::C_i$. We will assume every type to be a type class instance by using the special type class $\Type$ of all types.

Terms are formed from variables, typed constants, application, and lambda abstraction. Constants may be polymorphic in the sense that their types may contain free type variables. When a polymorphic constant is used, Isabelle automatically infers the type class instances for which the constant is used. Propositions are formed from implication, universal quantification over any type, and equality on any type.

We always assume that all types are fully reconstructed. Similarly, we cover neither the Isar proof language nor tactic invocations. Instead, we simply assume primitive inferences from Pure's natural deduction calculus \cite{isabellemanual}, i.e., using introduction/elimination rules for conjunction and implication, reflexivity and substitution rules for equality, as well as axioms for $\alpha\beta\eta$-conversion and extensionality.

%% file: lf.tex
The non-modular declarations in an LF signature are \emph{kinded type family} symbols $a:K$ and \emph{typed constants} $c:A$. Both may carry definitions, e.g., $c:A=t$ introduces $c$ as an abbreviations for $t$. The objects of Twelf are \emph{kinds} $K$, \emph{kinded type families} $A:K$, and \emph{typed terms} $t:A$. $\type$ is the kind of types, and $A\arr\type$ is the kind of type families indexed by terms of type $A$. We use Twelf notation for binding and application: The type $\Pi_{x:A}B(x)$ of dependent functions taking $x:A$ to an element of $B(x)$ is written $\tPi[A]{x}B\;x$, and the function term $\lambda_{x:A}t(x)$ taking $x:A$ to $t(x)$ is written $\tlam[A]{x}t\;x$. We write $A\arr B$ instead of $\tPi[A]{x}B$ if $x$ does not occur in $B$, and we will also omit the types of bound variables if they can be inferred.

The Twelf \emph{module system} \cite{rabeEA:twelfmod:09} is based on the notions of signatures and signature morphisms \cite{lfencodings}. Given two signatures $\itsig{S}{\Sigma}$ and $\itsig{T}{\Sigma'}$, a signature morphism from $S$ to $T$ is a type/kind-preserving map $\mu$ of $\Sigma$-symbols to $\Sigma'$-expressions. Thus, $\mu$ maps every constant $c:A$ of $\Sigma$ to a term $\mu(c):\ov{\mu}(A)$ and every type family symbol $a:K$ to a type family $\mu(a):\mu(K)$.
Here, $\mu(-)$ doubles as the homomorphic extension of $\mu$, which maps closed $\Sigma$-expressions to closed $\Sigma'$ expressions. Signature morphisms preserve typing and kinding, i.e., if $\vdash_\Sigma E:F$, then $\vdash_{\Sigma'}\mu(E):\mu(F)$.

\emph{Signature} declarations are straightforward: $\itsig{T}{\Sigma}$. Signatures may be nested and may include other signatures.
Basic \emph{morphisms} are given explicitly as $\{\sigma:S\arr T\}$, and composed morphisms are formed from basic morphisms, identity, composition, and two kinds of named morphisms: views and structures.
\footnote{Explicit morphisms are actually not present in \cite{rabeEA:twelfmod:09}. They are easy to add conceptually, but are a bit harder to add to Twelf as they violate the phase distinction between modular and non-modular syntax kept by the other declarations. We will need them later on.}

We will use the following grammar where the structure identifiers $\mpath{T}{s}$ and the symbol identifiers $\mc{S}{c}{\mu}$ and $\mc{S}{a}{\mu}$are described below:
\[\begin{array}{llll}
\mbox{Signature graphs}  & G    & \bnfas & \cdot \bnfalt G,\;\itsig{T}{\Sigma} \bnfalt G,\;\itviewt{v}{S}{T}{\mu} \\
\mbox{Signatures}       &\Sigma& \bnfas & \cdot \bnfalt \Sigma,\;\itsig{T}{\Sigma} \bnfalt  \Sigma,\;\itinclude{S}{} \\
                            &&&   \bnfalt \Sigma,\;\itstruct[\sigma]{s}{S}
                                 \bnfalt \Sigma,\;c:A[=t] \bnfalt  \Sigma,\;a:K[=A] \\
\mbox{Morphisms}        &\sigma&\bnfas & \cdot \bnfalt \sigma,\;\lfkw{struct}\;s:=\mu \hspace{.8cm}
                                               \bnfalt \sigma,\;c:=t \hspace{.5cm}\bnfalt \sigma,\;a:=A \\
\mbox{Compositions}     & \mu & \bnfas & \mpath{T}{s} \bnfalt \{\sigma:S\arr T\}\bnfalt  v 
                                           \bnfalt \mmtid \bnfalt \mmtincl \bnfalt \mu\;\mu\\
\mbox{Contexts}         & \Gamma& \bnfas & \cdot \bnfalt \Gamma,\;x:A \\
\mbox{Kinds}            & K   & \bnfas & \type \bnfalt  A \to K \\
\mbox{Type families}    & A   & \bnfas & \mc{S}{a}{\mu} \bnfalt A\;t \bnfalt \tPi[A]{x} A \\
\mbox{Terms}            & t   & \bnfas & \mc{S}{c}{\mu} \bnfalt x \bnfalt \tlam[A]{x} t \bnfalt t\;t \\
\end{array}\]

Modular LF uses the following judgments for well-formed syntax:
\begin{center}
\begin{tabular}{|l|l|}
\hline
$\isgraph{G}$             & well-formed signature graphs \\
$\omorphism{G}{\mu}{S}{T}$& morphism between signatures $S$ and $T$ declared in $G$ \\
$\iscont{G}{T}{\Gamma}$   & contexts for signature $T$ \\
$\otermtype{G;\Gamma}{T}{E}{E'}$ & $E$ has type/kind $E'$ over signature $T$ and context $\Gamma$\\
\hline
\end{tabular}
\end{center}

The judgment for signature graphs mainly formalizes uniqueness of identifiers and type-preservation of morphisms based on the typing judgment for expressions. The judgments for contexts and typing are essentially the same as for non-modular LF except that the identifiers available in signature $T$ and their types are determined by the module system. Therefore, we only describe the judgments for identifiers and morphisms and refer to \cite{RS:twelfmod:09} for details.

\paragraph{Morphisms}
In this paper, we only consider a simplified language and employ the following condition on all morphisms from $S$ to $T$: $T$ must include all signatures that $S$ includes, and if $S$ includes $R$, the application of $\mu$ to symbols of $R$ is the identity. In particular, views and structures may only be declared if this condition holds.\footnote{The Twelf implementation covers the general case.}

Firstly, the semantics of a \emph{structure} declaration $\itstruct[\sigma]{s}{S}$ in $T$ is that it is equivalent to the following \emph{induced declarations}: (i) for every constant $c:A$ of $S$ a constant $s.c:\mpath{T}{s}(A)$ in $T$, and (ii) a morphism $\mpath{T}{s}$ from $S$ to $T$ that maps every symbol $c$ of $S$ to $s.c$.
Here $\sigma$ is a partial morphism from $S$ to $T$, and if $\sigma$ contains $c:=t$, the constant $s.c$ is defined as $t$. In particular, $t$ must have type $\mpath{T}{s}(A)$ over $T$. The same holds for type family symbols $a$. Thus, structures instantiate parametric signatures.

\begin{wrapfigure}{r}{2.5cm}
\vspace{-1em}
\begin{tikzpicture}
\node (R) at (0,2) {$R$};
\node (S) at (0,0) {$S$};
\node (T) at (2,1) {$T$};
\draw[-\arrowtip](R) --node[right]{$\mpath{S}{r}$} (S);
\draw[-\arrowtip](S) --node[above]{$\mpath{T}{s}$} (T);
\draw[-\arrowtip](R) --node[above]{$\mu$} (T);
\end{tikzpicture}
\vspace{-2em}
\end{wrapfigure}

Because structures are named, a signature may have multiple structures of the same signature, which are all distinct. For example, if $S$ already contains a structure $r$ instantiating a third signature $R$, then $\itstruct{r'}{R}$ in $T$ leads to the two morphisms $r'$ and the composition $\mpath{S}{r}\;\mpath{T}{s}$ from $R$ to $T$ and two copies of the constants of $R$. Structures may instantiate whole structures at once: If $T$ declares instead $\itstruct[\lfkw{struct}\;r:=\mpath{T}{r'}]{r'}{R}$, then the two copies of $R$ are shared. More generally, $\sigma$ may contain instantiations $\lfkw{struct}\;r:=\mu$ for a morphism $\mu$ from $R$ to $T$, which is equivalent to instantiating every symbol $c$ of $R$ with $\mu(c)$. Another way to say this is that the diagram on the right commutes.

Secondly, the semantics of \emph{anonymous morphisms} $\{\sigma:S\arr T\}$ is straightforward. They are well-formed if $\sigma$ is total and map all constants according to $\sigma$. Thirdly, \emph{views} $v$ are just names given to existing morphisms.

Fourthly, \emph{inclusion}, \emph{identity} and \emph{composition} are defined by
\[\mathll{
 \ibnc{\itsig{T}{\Sigma}\minn G}
      {\itinclude{S}{}\minn \Sigma}
      {\omorphism{G}{\mmtincl}{S}{T}}
      {incl}
\nl
\ianc{\itsig{T}{\Sigma}\minn G}{\omorphism{G}{\mmtid}{T}{T}}{id}
\tb\tb
\ibnc{\omorphism{G}{\mu}{R}{S}}
     {\omorphism{G}{\mu'}{S}{T}}
     {\omorphism{G}{\mu\;\mu'}{R}{T}}
     {comp}
}\]
\medskip

\paragraph{Identifiers}
Defining which symbol identifiers are available in a signature is intuitively easy, but a formal definition can be cumbersome because all included symbols and those induced by structures have to be computed along with their translated types and definitions. Using morphisms and the novel notation $\mc{S}{c}{\mu}$ for symbol identifiers, we can give a very elegant definition:
 \[\icnc{\itsig{S}{\Sigma}\minn G}
        {c:E \minn \Sigma}{\omorphism{G}{\mu}{S}{T}}
        {\otermtype{G}{T}{\mc{S}{c}{\mu}}{\mu(E)}}
        {tp}\]
and similarly for defined symbols and type family constants. The price to pay is an awkward notation, but we can recover the usual notations as follows:
\begin{itemize}
	  \item $\mmtid$ yields local symbols, and we write $c$ instead of $\mc{T}{c}{\mmtid}$.
	  \item $\mmtincl$ yields included symbols, and we write $\mpath{S}{c}$ instead $\mc{S}{c}{\mmtincl}$.
	  \item If $T$ contains a structure from $S$, we have $\omorphism{G}{\mpath{T}{s}}{S}{T}$, and we write $s.c$ instead of $\mc{S}{c}{\mpath{T}{s}}$. Accordingly, we introduce constants $s.r.c$ for composed morphisms $\mpath{S}{r}\;\mpath{T}{s}$ from $R$ to $T$, and so on.
	  \item All other identifiers $\mc{T}{c}{\mu}$, e.g., those where $\mu$ contains views or anonymous morphisms, are reduced to one of the other cases by applying the morphism.
\end{itemize}

\paragraph{Functors}
While views are well-established in logical frameworks based on model theory (see, e.g., \cite{institutions,asl}), they are an unusual feature in proof theoretical frameworks. (In fact, the LF module system has been criticized for using views instead of functors or even -- in light of \cite{HarperPierce04} -- for using either one rather than only structures.) Therefore, we quickly describe how functors are a derived notion in the presence of views and anonymous morphisms.

Assume a functor $F$ from $S$ to $T$. Its input is a structure $s$ instantiating $S$, and its output is a list $\tau$ of instantiations for the symbols of $T$. Here $\tau$ may refer to the symbols induced by $s$. We can write this in LF as
\[\itsig{F_0}{\itstruct{s}{S}}\tb\itview{F}{T}{F_0}{\tau}\]
Now given a theory $D$, we can understand instances of a signature $S$ over $D$ as morphisms from $S$ to $D$. This is justified because a morphism from $S$ to $D$ realizes every declaration of $S$ in terms of $D$. More generally we can think of morphisms from $S$ to $D$ as implementations or models of $S$ in terms of $D$. The application of $F$ should map instances of $S$ to instances of $T$. Thus, given a morphism $\mu$ from $S$ to $D$, we can write the application $F(\mu)$ as the composed morphism
\[F\;\{\iassigs{s}{\mu}:F_0\arr D\}\]
which is indeed a morphism from $T$ to $D$.

%% file: classes.tex
We add a feature to the LF module system that permits morphism variables and abstraction over them. Therefore, we add the following productions to the grammar:
\[\begin{array}{llll@{\tb\tb}llll}
\mbox{Contexts}         & \Gamma& \bnfas & \Gamma,\;X:S &
\mbox{Compositions}     & \mu   & \bnfas & X  \\
\end{array}\]
Due to the presence of morphism variables, the judgment for well-formed morphisms must be amended to depend on the context. Then we can give the typing rules as:
\[\ibnc{\iscont{G}{T}{\Gamma}}{\itsig{S}{\Sigma}\minn G}{\iscont{G}{T}{\Gamma,\;X:S}}{contmor}
\tb\tb
\ianc{\iscont{G}{T}{\Gamma,\;X:S}}{\omorphism[\Gamma,X:S]{G}{X}{S}{T}}{morvar}\]
where we retain the restriction on signature inclusions: All signatures included into $S$ must also be included into $T$.

Note that we can understand the signature $S$ as a (dependent) record type, a morphism $\mu:S\arr T$ as a record value of type $S$ visible in the signature $T$, and an identifier $\mc{S}{c}{\mu}$ as the projection out of the record type $S$ at the field $c$ applied to $\mu$. Then $X$ is simply a variable of record type, and abstraction over morphism variables is straightforward:
\[\begin{array}{llll@{\tb\tb}llll}
\mbox{Type families}    & A     & \bnfas & \tPis[S]{X} A & 
\mbox{Terms}            & t     & \bnfas & \tlams[S]{X}t \bnfalt t\;\mu \\
\end{array}
\]
and (omitting the obvious $\Pi$-rule and the rules for $\beta$ and $\eta$-conversion)
\[
\ianc{\otermtype{G;\;\Gamma,\;X:S}{T}{t}{A}}
     {\otermtype{G;\;\Gamma}{T}{\tlams[S]{X}t}{\tPis[S]{X}A}}
     {morlam}
\tb\tb
\ibnc{\otermtype{G;\;\Gamma}{T}{f}{\tPis[S]{X}A}}
     {\omorphism[\Gamma]{G}{\mu}{S}{T}}
     {\otermtype{G;\Gamma}{T}{f\;\mu}{A[X/\mu]}}
     {morapp}
\]
Here $t$ and $A$ may contain occurrences of the morphism variable $X$. In particular $X$ may occur as a morphism argument to some expression, e.g., $g\;X$, or in an identifier $\mc{S}{c}{X}$, which we write as $X.c$ in accordance with our notation for structures.
\medskip

A crucial feature of the LF module system is that it is conservative: Modular signatures can be elaborated into non-modular ones (essentially by replacing every structure declaration with the induced constant declarations).
We want to elaborate morphism variables similarly.

To elaborate $X:S$, we can assume that all structures in $S$ have already been elaborated and that all defined symbols have been removed by expanding definitions, i.e., (up to reordering) $S$ is of the form $\itsig{S}{\itinclude{R_1}{},\ldots,\itinclude{R_m}{},\;c_1:B_1,\ldots, c_n:B_n}$. Then $\tlams[S]{X}t$ in a signature $T$ is elaborated to $\tlam[B_1']{x_1}\ldots\tlam[B_n']{x_n}t'$ where for expressions $E$ over $S$, we obtain $E'$ by replacing every occurrence of $X$ with the morphism $\{c_1:=x_1,\ldots,c_n:=x_n:S\arr T\}$. (In particular, after using morphism application, the identifiers $\mc{S}{c_i}{X}$ simply become $x_i$.) $\tPis[S]{X}A$ is elaborated accordingly. Finally, $t\;\mu$ is elaborated to $t\;\mu(\mpath{S}{c_1})\;\ldots\;\mu(\mpath{S}{c_n})$.

This extended module system is not conservative over LF: $S$ may contain type declarations, but LF does not permit abstraction over type variables. But we obtain conservativity if we make the following additional restriction: Contexts $\Gamma,X:S$ are only well-formed if all type family symbols $\mc{R}{a}{\mu}$ available in $S$ are included from other signatures, i.e., $\mu=\mmtincl \;\ldots \; \mmtincl$. Conversely, neither $S$ nor any signature that $S$ instantiates may contain type family declarations.

This restriction may appear to be introduced ad hoc, but in fact we consider it quite natural. Assume we have LF signatures $\ttlf{L}$ and $\ttlf{T}$ that represent an object logic $L$ and a theory $T$ of $L$. Then, typically, $\ttlf{L}$ contains type declarations for the syntactic categories and judgments of $L$ and constant declarations for the logical symbols and inference rules; $\ttlf{T}$ includes $\ttlf{L}$ and adds constant declarations for the non-logical symbols (sorts, functions, predicates, etc.) and axioms. Thus, our extension lets us abstract over morphisms out of theories but not over morphisms out of object logics. And the former are exactly the morphisms that we are interested in because morphisms out of $\ttlf{T}$ can be used to represent models or implementations of $T$. In particular, below, $T$ will be an axiomatic type class and morphisms out of $\ttlf{T}$ will be type class instances.

%% file: isa_lf.tex
The representation of Isabelle in LF proceeds in two steps. In a first step, we declare an LF signature $\Pure$ for the inner syntax of Isabelle. This syntax declares symbols for all primitives that can occur (explicitly or implicitly) in $\Pure$ expressions. In a second step, every Isabelle expression $E$ is represented as an LF expression $\ttlf{E}$. Finally we have to justify the adequacy of the encoding.

\begin{figure}[htb]
\begin{center}
\begin{tabular}{@{\tb}lllr}
\multicolumn{4}{l}{\lfkw{sig} $\Pure = \{$} \\
  $\tp$    & : & $\type$. \\
  $\func$  & : & $\tp \arr \tp \arr \tp$.                     & \lfkw{infix} right 0 $\func$. \\
  $\tm$    & : & $\tp \arr \type$.                            & \lfkw{prefix} 0 $\tm$. \\
  $\lambda$& : & $(\tm\; A \arr \tm\; B) \arr \tm\; (A \func B)$. \\
  $@$      & : & $\tm\;(A \func B) \arr \tm\; A \arr \tm\; B$.& \lfkw{infix} left 1000 $@$.\\[.5cm]
  
  $\prop$  & : & $\tp$. \\
  $\bigwedge$& : &  $(\tm\; A \arr \tm\; \prop) \arr \tm\;\prop$. \\
  $\iimpl$ & : &  $\tm\;\prop \arr \tm\;\prop \arr \tm\;\prop$. & \lfkw{infix} right 1 $\iimpl$. \\
  $\equiv$ & : &  $\tm\;A \arr\tm\; A \arr \tm\;\prop$.       & \lfkw{infix} none 2 $\equiv$. \\[.5cm]
  
  $\vdash$     & : & $\tm\;\prop \arr \type$.                 & \lfkw{prefix} 0 $\vdash$. \\
  $\bigwedge$I & : & $({x: \tm\;A} \vdash (B\;x)) \;\arr\; \vdash \bigwedge ([x] B\;x)$. \\
  $\bigwedge$E & : & $\vdash \bigwedge ([x] B\; x) \;\arr\; \{x: \tm\;A\} \vdash (B\;x)$. \\
  $\iimpl$I    & : & $(\vdash A \;\arr\; \vdash B) \;\arr\; \vdash A \iimpl B$. \\
  $\iimpl$E    & : & $\vdash A \iimpl B \;\arr\; \vdash A \;\arr\; \vdash B$. \\
  refl         & : & $\vdash X \equiv X$. \\
  subs         & : & $\{F: \tm\; A \;\arr\; \tm\; B\} \vdash X \equiv Y \;\arr\; \vdash F\;X \equiv F\; Y.$ \\
  exten        & : & $\{x : \tm\;A\} \ided (F\;x) \equiv (G\;x) \;\arr\;\ided \lambda F \equiv\lambda G$. \\
  beta         & : & $\ided (\lambda [x: \tm\; A] F\;x)\; @\; X \equiv F\; X$. \\
  eta          & : & $\ided \lambda \;([x: \tm\; A] F \;@\; x) \equiv F$.  \\
  \multicolumn{4}{@{\tb}l}{$\itsig{\Type}{\tclass:\tp.}$.} \\
\multicolumn{4}{l}{$\}$.}
\end{tabular}
\caption{LF Signature for Isabelle}\label{fig:isa:isalf}
\end{center}
\end{figure}

For the \emph{inner syntax}, the LF signature $\Pure$ is given in Fig.~\ref{fig:isa:isalf}. This is a straightforward intrinsically typed encoding of higher-order logic in LF (e.g., as in \cite{lf}). Pure types $\tau$ are encoded as LF-terms $\ttlf{\tau}:\tp$ and Pure terms $t::\tau$ as LF-terms $\ttlf{t}:\tm\;\ttlf{\tau}$. Using higher-order abstract syntax, the LF function space $A\arr B$ with $\lambda$-abstraction $\tlam[A]{x}t$ and application $f\;t$ is distinguished from the encoding $\tm\;(\ttlf{\sigma}\func \ttlf{\tau})$ of the Isabelle function space with application $\ttlf{f}\,@\,\ttlf{t}$ and $\lambda$-abstraction $\lambda (\tlam[\tm\;\ttlf{\tau}]{x}\ttlf{t})$. Pure propositions $\phi$ are encoded as LF-terms $\ttlf{\phi}:\tm\;\prop$, and Pure inferences $P$ proving $\phi$ as LF-terms $\ttlf{P}$ of type $\ided\;\ttlf{\phi}$. Where possible, we use the same symbol names in LF as in Isabelle, and we can also mimic most of the Isabelle operator fixities and precedences.

The signature $\Pure$ only encodes how composed Pure expressions are formed from the atomic ones. The atomic expressions -- variables and constants etc. -- are added when encoding the outer syntax as LF declarations. For the \emph{non-modular declarations}, this is straightforward, an overview is given in the following table:
\begin{center}
\begin{tabular}{|l|l|l|}
\hline
Expression              & Isabelle                        & LF \\ \hline
base type, type operator& $(\alpha_1,\ldots,\alpha_n)\;t$ & $t:\tp\arr\ldots\arr\tp\arr\tp$ \\
type variable           & $\alpha$                        & $\alpha:\tp$ \\
constant                & $c::\tau$                       & $c:\tm\;\ttlf{\tau}$ \\
variable                & $x::\tau$                       & $x:\tm\;\ttlf{\tau}$ \\
assumption/axiom/definition & $a:\phi$                    & $a:\;\ided\ttlf{\phi}$ \\
theorem                 & $a:\phi\;P$                     & $a:\;\ided\ttlf{\phi}=\ttlf{P}$\\
\hline
\end{tabular}
\end{center}

The main novelty of our encoding is to also cover the \emph{modular declarations}. The basic idea is to represent all high-level scoping concepts as signatures and all relations between them as signature morphisms as in the following table:
\begin{center}\begin{tabular}{|l|l|}
\hline
Isabelle         & LF \\ \hline
theory, locale, type class           & signature \\
theory import                        & morphism (inclusion) \\
locale import, type class import     & morphism (structure) \\
sublocale, interpretation, type class instantiation    & morphism (view) \\
instance of type class $C$           & morphism with domain $C$ \\
\hline
\end{tabular}
\end{center}
{
\renewcommand{\meta}[1]{{\color{red}#1}}
In the following, we give the important cases of the mapping  $\ttlf{-}$ from Isabelle to LF by induction on the Isabelle syntax. We occasionally  use \meta{color} to distinguish the meta-level symbols (such as \meta{=}) from Isabelle and Twelf syntax such as $=$.

\paragraph{Theories}
Isabelle theories and theory imports are encoded directly as LF-signatures and signature inclusions. The only subtlety is that the LF encodings additionally include our $\Pure$ signature.
\begin{quote}
 \ttlf{\keyw{theory} $T$ \keyw{imports} $T_1,\ldots,T_n$ \keyw{begin} $\Sigma$ \keyw{end}} \meta{=} \\
 \lfkw{sig} $T=\{$ \lfkw{include} $\Pure$. \lfkw{include} $T_1$. \ldots \lfkw{include} $T_n$. \ttlf{$\Sigma$}$\}$.
\end{quote}
where the body $\Sigma$ of the theory is translated component-wise as described by the respective cases below.

\paragraph{Type Classes}
The basic idea of the representation of Isabelle type classes in LF is as follows: An Isabelle type class $C$ is represented as an LF signature $C$ that contains all the declarations of $C$ \emph{and} a field $\tclass:\tp$. All occurrences in $C$ of the single permitted type variable $\alpha::C$ are translated to $\tclass$ such that $\tclass$ represents the type that is an instance of $C$.

This means that $\alpha$ is not considered as a type variable but as a type declaration that is present in the type class. This change of perspective is essential to obtain an elegant encoding of type classes.

In particular, the subsignature $\Type$ of $\Pure$ represents the type class of all types. Morphisms with domain $\Type$ are simply terms of type $\tp$, i.e., types.

The central invariant of the representation is this: An Isabelle type class instance $\tau::C$ is represented as an LF morphism $\ttlf{\tau::C}$ from $C$ into the current LF signature that maps the field $\tclass$ to $\ttlf{\tau}$ and all operations of $C$ to the encoding of their definitions at $\tau$. Thus, in particular, $\ttlf{\tau::C}(C.\tclass)=\ttlf{\tau}$.

\begin{example}[Continued]
The first type class from Ex.~\ref{ex:isa:isa} is represented in LF as follows:
\[\begin{array}{l}
\itsig{order}{\tclass:\tp.\;\leq:\tm(\tclass\func\tclass\func\prop)}
\end{array}\]
\end{example}

In general, we represent type classes as follows:
 \begin{quote}
   \ttlf{\keyw{class} $C=C_1\;\ldots\;C_n+\Sigma$} \meta{=}
   \lfkw{sig} $C = \{\tclass:\tp.\;
                      I_1.\;\ldots\;I_n. \;
                      \ttlf{\Sigma}\}.$
 \end{quote}
where $I_i$ abbreviates $\itstruct[\iassig{\tclass}{\tclass}\;\rho_i]{ins_i}{C_i}$ for some fresh names $ins_i$.
Since one $\tclass$ is imported from each superclass $C_i$, they must be shared using the instantiations $\iassig{\tclass}{\tclass}$. $\rho_i$ contains one structure sharing declaration for each type class imported by $I_i$ that has already been imported by $I_1,\ldots,I_{i-1}$.

\begin{example}[Continued]
The second type class from Ex.~\ref{ex:isa:isa} is represented in LF as follows:
\[\begin{array}{l}
\itsig{semlat}{\tclass:\tp.\;
  \itstruct[\iassig{\tclass}{\tclass}]{o}{order}.\;
  \sqcap:\tm(\tclass\func\tclass\func\tclass)}
\end{array}\]
\end{example}

A type class \emph{instantiation}
 \[\keyw{instantiation}\;t :: (C_1,\ldots,C_n)C\; \keyw{begin}\;\Sigma\;\pi\; \keyw{end}\]
is represented as an LF functor taking instances of the $C_i$ and returning an instance of $C$. We represent such a functor as a signature
 \[\lfkw{sig}\;\nu = \{\itstruct{\alpha_1}{C_1} \ldots \itstruct{\alpha_n}{C_n}\}.\]
collecting the input and a view
 \[\lfkw{view}\;\nu': C\arr \nu = \{\iassig{\tclass}{t\;\alpha_1.\tclass\;\ldots\;\alpha_n.\tclass}\;\ttlf{\Sigma} \ttlf{\pi}\}.\]
describing the output. $\nu'$ must map the field $\tp$ of $C$ to the type that is an instance of $C$. This type is obtained by applying $t$ to the argument types that are instances of the $C_i$. In Isabelle, this is $t\;\alpha_1\;\ldots\;\alpha_n$; in LF, each $\alpha_i$ is a structure of $C_i$, thus we use the induced constants $\alpha_i.\tclass$.

Here $\ttlf{\Sigma}$ gives instantiations that map every constant of $C$ to its definition in terms of the $\alpha_i$. Similarly, $\ttlf{\pi}$ maps every axiom of $C$ to its proof. Note how -- in accordance with the Curry-Howard representation of proofs as terms -- the discharging of proof obligations is just a special case of instantiating a constant.

Now assume type class instances $\tau_i::C_i$ encoded as morphisms $\ttlf{\tau_i::C_i}:C_i\arr S$ (where $S$ is the current signature). The encoding $\ttlf{(\tau_1,\ldots,\tau_n)t::C}:C\arr S$ is obtained as the composition
\[\nu'\;\{\iassigs{\alpha_1}{\ttlf{\tau_1::C_1}}\;\ldots\;\iassigs{\alpha_n}{\ttlf{\tau_n::C_n}}:\nu\arr S\}.\]
Clearly this is a morphism from $C$ to $S$; we need to show that indeed \[\ttlf{(\tau_1,\ldots,\tau_n)t::C}(\mpath{C}{\tclass})\;\meta{=}\;\ttlf{t\;\tau_1\;\ldots\;\tau_n}.\]
This holds because
\[\mathll{
   \ttlf{(\tau_1,\ldots,\tau_n)t::C}(\mpath{C}{\tclass}) \;\meta{=}\;
   \{\ldots\iassigs{\alpha_i}{\ttlf{\tau_i::C_i}}\ldots\}(\nu'(\mpath{C}{\tclass}))\nl
   \{\ldots\iassigs{\alpha_i}{\ttlf{\tau_i::C_i}}\ldots\}
          (t\;\alpha_1.\tclass\;\ldots\;\alpha_n.\tclass) \;\meta{=}\;\nl
   t\;\ttlf{\tau_1::C_1}(\mpath{C_1}{\tclass})\;\ldots\;\ttlf{\tau_n::C_n}(\mpath{C_n}{\tclass})\;\meta{=}\;
   t\;\ttlf{\tau_1}\;\ldots\;\ttlf{\tau_n}\;\meta{=}\;\ttlf{t\;\tau_1\;\ldots\;\tau_n}
}\]
\medskip

We have the general result that the Isabelle subclass relation $C\sq D$ holds iff there is an LF morphism $i:D\arr C$. Then if the type class instance $\tau::C$ (occurring in some theory or locale $S$) is represented as a morphism $\ttlf{\tau::C}:C\arr S$, the type class instance $\tau::D$ is represented as $\ttlf{\tau::D}\;\meta{=}\;i\;\ttlf{\tau::C}$. Isabelle has the limitation that there can be at most one way how $C$ is a subclass of $D$, which has the advantage that $i$ is unique and can be dropped from the notation. In LF, we have to make it explicit.

\begin{example}[Continued]
The trivial subclass relation $order\sq \Type$ is represented by the morphism $i\;\meta{=}\;\{\iassig{\tclass}{\tclass}:\Type\arr order\}$. The subclass relation $semlat\sq order$ is represented by the morphism $\mpath{semlat}{o}$. Finally, the morphism $i\;\mpath{semlat}{o} \;\;\meta{=}\;\;\{\iassig{\tclass}{\tclass}:\Type\arr semlat\}$ represents $semlat\sq \Type$.
\end{example}

\paragraph{Locales}
Similarly to type classes, Isabelle \emph{locales} are encoded as subsignatures: For example,
  \[\keyw{locale}\; loc=ins_1:loc_1\; \keyw{where}\; \sigma_1\; \keyw{for}\; \Sigma + \Sigma'\]
is encoded as the LF signature
  \[\lfkw{sig}\;loc=\{\Theta\;\ttlf{\Sigma}\; \itstruct[\ttlf{\sigma_1}]{ins_1}{loc_1}.\;\ttlf{\Sigma'} \}.\]

Here $\Theta$ contains type declarations $\alpha:\tp$ for the free type variables of the locale. Those are the free type variables that occur in the declarations of $\Sigma$ and $\Sigma'$. These correspond to the single declaration $\tclass:\tp$ in type classes. This encoding of type variables may be surprising because free type variables correspond to universal types whereas the declarations in $\Theta$ correspond to existential types. We hold that our LF-encoding precisely captures the intended meaning of locales, whereas the definition of locales within Isabelle prefers universal type variables in order to be compatible with the underlying type theory.

If a locale inherits from more than one locale, the encoding is defined correspondingly using one structure $\itstruct[\theta_i\;\ttlf{\sigma_i}]{ins_i}{loc_i}$ for each locale instance $ins_i:loc_i\; \keyw{where} \;\sigma_i$.
Here $\theta_i$ contains the instantiations for free type variables of $loc_i$ that are induced by $\sigma_i$ and inferred by Isabelle.
Furthermore, some additional sharing declarations become necessary due to a subtlety in the semantics of Isabelle locales: If a locale inherits two equal instances (same locale, same instantiations), they are implicitly identified. But in LF different structures are always distinguished unless shared explicitly.
Therefore, we have to add to $\ttlf{\sigma_i}$ one sharing declaration $\lfkw{struct}\;ins:=ins'$ for each instance $ins$ present in $loc_i$ that is equal to one already imported by one of $ins_1$,\ldots, $ins_{i-1}$.

\begin{example}[Continued]
The locale from Ex.~\ref{ex:isa:isa} is represented in LF as follows:
\[\begin{array}{l}
\lfkw{sig}\;lat=\{ \\
 \tb \itstruct{inf}{semlat}.\\
 \tb \itstruct[
    \iassig{\tclass}{inf.\tclass}.\;\iassig{\leq}{\lambda \tlam{x}\lambda\tlam{y}\;\inf.\leq\,@\,y@\,x}]
    {sup}{semlat}. \\
\}
\end{array}\]
Note how the instantiation for $\leq$ induces an instantiation for the type $\tclass$. In other words, the $\theta$ mentioned above is $\iassig{\tclass}{inf.\tclass}$.
\end{example}

\emph{Sublocale} declarations are encoded as views from the super- to the sublocale. Thus, the declaration
  \[\keyw{sublocale}\; loc'<loc \;\keyw{where}\;\sigma\;\pi\]
  is encoded as (for some fresh name $\nu$):
  \[\lfkw{view}\;\nu:loc\arr loc'=\{\theta\;\ttlf{\sigma}\;\ttlf{\pi}\}.\]
Here $\theta$ contains the instantiations of the free type variables (see $\Theta$ above), which are inferred by Isabelle based on the instantiations in $\sigma$.

Locale \emph{interpretations} are interpreted in the same way except that the codomain is the current LF signature (which encodes the Isabelle theory containing the locale interpretation) instead of the sublocale.

As for type classes, we have the general result that $loc$ is a sublocale of $loc'$ iff there is an LF signature morphism from $loc$ to $loc'$. Accordingly, $loc$ can be interpreted in the theory $T$ iff there is a morphism from $loc$ to $T$. For example, $loc$ is a sublocale of $loc_1$ from above via the composed morphism $\nu\;\mpath{loc}{ins_1}$. Contrary to type classes, there may be several different sublocale relationships between two locales. In LF these are distinguished elegantly as different morphisms between the locales.

\begin{example}[Continued]
$lat$ is a sublocale of $semlat$ in two different ways represented by the LF morphisms $\mpath{lat}{inf}$ and $\mpath{lat}{sup}$. These are trivial sublocale relations induced by inheritance.
\end{example}

\paragraph{Constant Declarations}
Finally we have to represent those aspects of the non-modular declarations that are affected by type classes. We will only consider the case of constants. Definitions, axioms, and theorems are represented accordingly. The central idea is that free type variables constrained by type classes are represented using $\lambda$ abstraction for morphism variables.

An Isabelle constant $c::\tau$ with free \emph{type variables} $\alpha_i::C_i$ is represented as the LF-constant taking morphism arguments:
\[c:\tPis[C_1]{\alpha_1}\ldots\tPis[C_n]{\alpha_n}\tm\;\ttlf{\tau}.\]
Here in $\ttlf{\tau}$ every occurrence of the morphism variable $\alpha_i$ is represented as $\alpha_i.\tclass$.

Whenever $c$ is used with inferred type arguments $\tau_i::C_i$ in a composed expression, it is represented by application of $c$ to morphisms: \[\ttlf{c}=c\;\ttlf{\tau_1::C_1}\;\ldots\;\ttlf{\tau_n::C_n}.\]
Actually, we cannot use the same identifier $c$ in LF as in Isabelle: Instead, we must keep track how $c$ came into scope. For example, if $c$ was imported from some theory $S$, we must use $S.c$ in LF; if the current scope is a locale and $c$ was imported from some other locale via an instance $ins$, we must use $ins.c$ in LF; if $c$ was moved into the current theory from a locale $loc$ via an \keyw{interpretation} declaration which was encoded using the fresh name $\nu$, we must use $\mc{loc}{c}{\nu}$ in LF, and so on.

\paragraph{Types}
The representation of types was already indicated above, but we summarize it here for clarity. Type operator declarations $(\alpha_1,\ldots,\alpha_n)t$ are encoded as constants $t:\tp\arr\ldots\arr\tp\arr\tp$. And types occurring in expressions are encoded as
\[\begin{array}{lcl}
 \ttlf{\alpha::C} & = & \alpha.\tclass \\
 \ttlf{t}         & = & t \\
 \ttlf{(\tau_1,\ldots,\tau_n)t} & = & t\;\ttlf{\tau_1}\;\ttlf{\tau_n} \\
 \ttlf{\tau_1\func\tau_n}     & = & \ttlf{\tau_1}\func\ttlf{\tau_2} \\
 \ttlf{\prop}     & = & \prop.
\end{array}\]
}

\paragraph{Adequacy}
Before we state the adequacy, we need to clarify in what sense our representation is adequate.
In Isabelle, locales and type classes are not primitive notions. Instead, they are internally elaborated into the underlying type theory. For example, all declarations in a locale or a type class are relativized and lifted to the top level. Thus, they are available elsewhere and not only within the locale.
While there are certainly situations when this is useful, here we care about the modular structure and the underlying type theory, but not about the elaboration of the former into the latter. Therefore, we do not want a representation in LF that adequately preserves the elaboration. In fact, if we wanted to preserve the elaboration, we could simply use Isabelle to eliminate all modular structure and represent the non-modular result using well-known representations of higher-order logic in LF.

Therefore, we have to forbid all Isabelle theories where names are used outside their scope. Let us call an Isabelle theory \emph{simple} if all declared names are only used in their respective declaration scope -- theory, locale, or type class -- unless they were explicitly moved into a new scope using \keyw{imports}, \keyw{sublocale}, \keyw{interpretation}, or \keyw{instantiation} declarations, or using inheritance between type classes and locales.

Then we can summarize our representation with the following theorem:
\begin{theorem}
A simple sequence of Isabelle theories $T_1\;\ldots\;T_n$ is well-formed (in the sense of Isabelle) iff the LF signature graph $\Pure\;\ttlf{T_1}\;\ldots\;\ttlf{T_n}$ is well-formed (in the sense of LF extended with morphism variables).
\end{theorem}
\begin{proof}
To show the adequacy for the encoding of the inner syntax is straightforward. A similar proof was given in \cite{lf}.

The major lemmas for the outer syntax were already indicated in the text:
\begin{itemize}
	\item For an Isabelle type class instance $\tau::C$ used in theory or locale $S$ and context $\Gamma$, we have $\omorphism[\Gamma]{G}{\ttlf{\tau::C}}{C}{S}$ and $\ttlf{\tau::C}(C.\tclass)=\ttlf{\tau}$.
	\item There is an Isabelle sublocale relation $loc'<loc$ via instantiations $\sigma$ whenever the incomplete LF morphism $\{\ttlf{\sigma}\;\ldots : loc\arr loc'\}$ can be completed (by instantiating the axioms of $loc$ with proof terms over $loc'$).
\end{itemize}

The main difficulty in the proofs is to show that at any point in the translated LF signatures exactly the right atomic expressions are in scope. This has to be verified by a difficult and tedious comparison of the Isabelle documentation with the semantics of the LF module system. In particular, in our simplified grammar for Isabelle, we have omitted the features that would break this result. These include in particular the features whose translation requires inventing and keeping track of fresh names, such as overloading and unqualified locale instantiation.
\end{proof}
The above proof is not quite convincing, even vague. The problem is that a more elaborate proof would require formal definitions of well-formedness for both module systems, and these are beyond the scope of this paper. (In fact, no comprehensive reference definition is available yet for the semantics of the modular syntax of either system.)

%% file: conc.tex
We have presented a representation of Isabelle's module system in the LF module system. Previous logic encodings in LF have only covered non-modular languages (e.g., \cite{lf,lfencodings,lfcut}), and ours is the first encoding of a modular logic. We also believe ours to be the first encoding of type classes or locale-like features in any logical framework.

The details of the translation are quite difficult, and a full formalization requires intricate knowledge of both systems. However, guided by the use of signatures and signature morphisms as the main primitives in the LF module system, we could give a relatively intuitive account of Isabelle's structuring mechanisms.

Our translation preserves modular structure; in particular the translation is compositional and the size of the output is linear in the size of the input. We are confident that our approach scales to other systems such as the type classes of Haskell or the functors of SML, and thus lets us study the modular properties of programming languages in logical frameworks. Moreover, we hold that the trade-off made in the LF module system between expressivity and simplicity makes it a promising starting point to investigate the movement of modular developments between systems.

In order to formulate the representation, we had to add abstraction over morphisms to the LF module system. This effectively gives LF a restricted version of dependent record types. This is similar to the use of contexts as dependent records as, e.g., in \cite{beluga}. Contrary to, e.g., \cite{nuprl} and \cite{agda}, the LF records may only occur in contravariant positions, which makes them a relatively simple conservative addition.

An integration of this feature into the Twelf implementation of LF remains future work. Similarly, the use of anonymous morphisms has not been implemented in Twelf yet. In both cases, the implementation is conceptually straightforward. However, since it would permit the use of morphisms in terms, types, and kinds, it would require a closer integration of modular and core syntax in Twelf, which has so far been avoided deliberately. We will undertake the Twelf side of the implementation soon.

In any case, Twelf will hardly be a bottleneck. Any implementation of a translation from Isabelle to LF would have to be implemented from within Isabelle as it requires Isabelle's reconstruction of types and instantiations (let alone proof terms). However, Isabelle currently eliminates most aspects of modularity when checking a theory. For example, it is already difficult to export the local constants of a theory because the methods provided by Isabelle can only return all local, imported, or internally generated constants at once. The most promising albeit still very difficult approach seems to be to use a standalone parser for the Isabelle outer syntax and then fill in the gaps by calling the methods provided by Isabelle. Thus, even though this paper solves the logical questions of how to translate from Isabelle to LF, the corresponding software engineering questions are non-trivial and remain open.

%% file: paper.bbl
\providecommand\seen{seen } \providecommand\webpageat{web page at }
  \providecommand\homepageat{home page at }
  \providecommand\projectpageat{project page at }
  \providecommand\systempageat{system home page at }
  \providecommand\svnrepoat{Subversion repository at }
  \providecommand\January{January} \providecommand\February{February}
  \providecommand\March{March} \providecommand\April{April}
  \providecommand\May{May} \providecommand\June{June}
  \providecommand\July{July} \providecommand\August{August}
  \providecommand\September{September} \providecommand\October{October}
  \providecommand\November{November} \providecommand\December{December}
  \providecommand\AUSTRALIA{Australia} \providecommand\ROMANIA{Romania}
  \providecommand\MEXICO{Mexico} \providecommand\ITALY{Italy}
  \providecommand\USA{USA} \providecommand\IRELAND{Ireland}
  \providecommand\HUNGARY{Hungary} \providecommand\JAPAN{Japan}
  \providecommand\CANADA{Canada} \providecommand\SPAIN{Spain}
  \providecommand\NETHERLANDS{Netherlands} \providecommand\UK{UK}
  \providecommand\SWEDEN{Sweden} \providecommand\GERMANY{Germany}
  \providecommand\openmath{OpenMath} \providecommand\fc{forthcoming}
  \providecommand\PROC{Proceedings} \providecommand\omdoc{OMDoc}
  \providecommand\activemath{ActiveMath}

%% file: paper.bbl
\begin{thebibliography}{10}

\bibitem{lambdacube}
H.~Barendregt.
\newblock Lambda calculi with types.
\newblock In S.~Abramsky, D.~Gabbay, and T.~Maibaum, editors, {\em {Handbook of
  Logic in Computer Science}}, volume~2. Oxford University Press, 1992.

\bibitem{churchtypes}
A.~Church.
\newblock {A Formulation of the Simple Theory of Types}.
\newblock {\em {Journal of Symbolic Logic}}, 5(1):56--68, 1940.

\bibitem{nuprl}
R.~Constable, S.~Allen, H.~Bromley, W.~Cleaveland, J.~Cremer, R.~Harper,
  D.~Howe, T.~Knoblock, N.~Mendler, P.~Panangaden, J.~Sasaki, and S.~Smith.
\newblock {\em {Implementing Mathematics with the Nuprl Development System}}.
\newblock Prentice-Hall, 1986.

\bibitem{institutions}
J.~Goguen and R.~Burstall.
\newblock Institutions: Abstract model theory for specification and
  programming.
\newblock {\em Journal of the Association for Computing Machinery},
  39(1):95--146, 1992.

\bibitem{isabelle_classes}
F.~Haftmann and M.~Wenzel.
\newblock {Constructive Type Classes in Isabelle}.
\newblock In T.~Altenkirch and C.~McBride, editors, {\em TYPES conference},
  pages 160--174. Springer, 2006.

\bibitem{lf}
R.~Harper, F.~Honsell, and G.~Plotkin.
\newblock {A framework for defining logics}.
\newblock {\em {Journal of the Association for Computing Machinery}},
  40(1):143--184, 1993.

\bibitem{HarperPierce04}
R.~Harper and B.~Pierce.
\newblock {Design Issues in Advanced Module Systems}.
\newblock In B.~Pierce, editor, {\em {Advanced Topics in Types and Programming
  Languages}}. {MIT Press}, 2005.

\bibitem{lfencodings}
R.~Harper, D.~Sannella, and A.~Tarlecki.
\newblock Structured presentations and logic representations.
\newblock {\em Annals of Pure and Applied Logic}, 67:113--160, 1994.

\bibitem{isabelle_locales}
F.~Kamm{\"u}ller, M.~Wenzel, and L.~Paulson.
\newblock {Locales -- a Sectioning Concept for Isabelle}.
\newblock In Y.~Bertot, G.~Dowek, A.~Hirschowitz, C.~Paulin, and L.~Thery,
  editors, {\em {Theorem Proving in Higher Order Logics}}, pages 149--166.
  Springer, 1999.

\bibitem{martinlof}
P.~Martin-{L\"o}f.
\newblock {An Intuitionistic Theory of Types: Predicative Part}.
\newblock In {\em {Proceedings of the '73 Logic Colloquium}}, pages 73--118.
  North-Holland, 1974.

\bibitem{isar}
T.~Nipkow.
\newblock {Structured Proofs in Isar/HOL}.
\newblock In H.~Geuvers and F.~Wiedijk, editors, {\em TYPES conference}, pages
  259--278. Springer, 2002.

\bibitem{agda}
U.~Norell.
\newblock {The Agda WiKi}, 2005.
\newblock \url{http://wiki.portal.chalmers.se/agda}.

\bibitem{isabelle1}
L.~Paulson.
\newblock {The Foundation of a Generic Theorem Prover}.
\newblock {\em {Journal of Automated Reasoning}}, 5(3):363--397, 1989.

\bibitem{isabelle}
L.~Paulson.
\newblock {\em {Isabelle: A Generic Theorem Prover}}, volume 828 of {\em
  Lecture Notes in Computer Science}.
\newblock Springer, 1994.

\bibitem{lfcut}
F.~Pfenning.
\newblock Structural cut elimination: I. intuitionistic and classical logic.
\newblock {\em Information and Computation}, 157(1-2):84--141, 2000.

\bibitem{twelf}
F.~Pfenning and C.~Sch{\"u}rmann.
\newblock System description: {Twelf} - a meta-logical framework for deductive
  systems.
\newblock {\em Lecture Notes in Computer Science}, 1632:202--206, 1999.

\bibitem{beluga}
B.~Pientka and J.~Dunfield.
\newblock {A Framework for Programming and Reasoning with Deductive Systems
  (System description)}.
\newblock In {\em {International Joint Conference on Automated Reasoning}},
  2010.
\newblock To appear.

\bibitem{RS:twelfmod:09}
F.~Rabe and C.~Sch{\"u}rmann.
\newblock {A Practical Module System for LF}.
\newblock In J.~Cheney and A.~Felty, editors, {\em {Proceedings of the Workshop
  on Logical Frameworks: Meta-Theory and Practice (LFMTP)}}, pages 40--48. ACM
  Press, 2009.

\bibitem{rabeEA:twelfmod:09}
F.~Rabe and C.~Sch{\"u}rmann.
\newblock A practical module system for {LF}.
\newblock In {\em {Proceedings of the Workshop on Logical Frameworks
  Meta-Theory and Practice (LFMTP)}}, 2009.

\bibitem{asl}
D.~Sannella and M.~Wirsing.
\newblock {A Kernel Language for Algebraic Specification and Implementation}.
\newblock In M.~Karpinski, editor, {\em {Fundamentals of Computation Theory}},
  pages 413--427. Springer, 1983.

\bibitem{isabellemanual}
M.~Wenzel.
\newblock {The Isabelle/Isar Reference Manual}, 2009.
\newblock \url{http://isabelle.in.tum.de/documentation.html}, Dec 3, 2009.

\end{thebibliography}
